# Closing the gap between research and projects in climate change innovation in Europe

Francesca Larosa[1,5*], Jaroslav Mysiak[5,6], Marco Molinari[2], Panagiotis Varelas[8,9], Haluk Akay[1,2], Will McDowall[3], Catalina Spadaru[4], Francesco Fuso-Nerini[1,2], Ricardo Vinuesa[7,1]

**Teaser:** Publicly funded projects in climate change innovation in Europe require better connections to academic research

**Abstract**

Innovation is a key component to equip our society with tools to adapt to new climatic conditions. The development of research-action interfaces shifts useful ideas into operationalized knowledge allowing innovation to flourish. In this paper we quantify the existing gap between climate research and innovation action in Europe using a novel framework that combines artificial intelligence (AI) methods and network science. We compute the distance between key topics of research interest from peer review publications and core issues tackled by innovation projects funded by the most recent European framework programmes. Our findings reveal significant differences exist between and within the two layers. Economic incentives, agricultural and industrial processes are differently connected to adaptation and mitigation priorities. We also find a loose research-action connection in bioproducts, biotechnologies and risk assessment practices, where applications are still too few compared to the research insights. Our analysis supports policy-makers to measure and track how research funding result in innovation action, and to adjust decisions if stated priorities are not achieved.

**Towards an innovation-led and sustainable society**

Climate change is an existential threat to the productive means of societies, but also to the livelihoods of human beings, their cultural heritage, behaviors and habits. The inherent physical complexity of the climate, together with the pressure of anthropogenic activities, require long-term and comprehensive solutions capable of accounting for multiple factors. According to the International panel on Climate Change (IPCC), *"innovation and change can expand the availability and/or*

[1] KTH Climate Action Centre, Royal Institute of Technology (KTH), Stockholm, Sweden
[2] Division of Energy Systems, Royal Institute of Technology (KTH), Stockholm, Sweden
[3] Institute for Sustainable Resources, University College London (UCL), London, United Kingdom
[4] Energy Institute, University College London (UCL), London, United Kingdom
[5] Euro-Mediterranean Center on Climate Change (CMCC), Venice, Italy
[6] Ca' Foscari University, Venice, Italy
[7] FLOW, Engineering Mechanics, KTH Royal Institute of Technology, Stockholm, Sweden
[8] Aero-Thermo-Mechanics Laboratory, École Polytechnique de Bruxelles, Université Libre de Bruxelles, Belgium
[9] Institute of Mechanics, Materials, and Civil Engineering, Université Catholique de Louvain, Belgium
*Corresponding author:* larosa@kth.se

*effectiveness of adaptation and mitigation options"*[1]. Innovation acts as an interface between research, technological development, and industrial policy[2]. By offering technical solutions[3], as well as new or improved management practices, innovation triggers policy [4] and socio-cultural changes [5] and help societies charting the complexities of a changing climate.

The economics of technological innovation has become a prominent research interest over the last decade[6]. Literature has devoted a wide attention to knowledge production and diffusion processes, shifting from a linear to a co-production understanding of science-policy interfaces[7]. Applied and theoretical research is part of these interfaces[8] and help creating communities which populate innovation pathways[9]. When it comes to climate change, the assessment of a healthy research-innovation landscape contributes to the transformation of the core structures of the system[10]: science informs climate risk reduction strategies and contributes to the creation of solutions to reduce emissions. Research has acknowledged the relevance of science in stimulating innovation[11] especially in climate change research[12]. Furthermore, there is growing awareness of the need for new methods to assess and evaluate research outputs[13], as institutions are moving fast towards evidence-based approaches to help solving major societal challenges. Despite growing attention to publicly funded research and development (R&D) programmes and to their interaction with the existing knowledge stock[14], there is still a critical gap in making evaluation processes efficient and scalable while also accounting for spillovers, trade-offs and synergies[15], and how priorities identified by research reflect in concrete innovation actions and incentives.

To fill in the gaps, this research explores how the landscape of the European innovation system for climate change interact with research. As the European Commission has confirmed the launch of a Union-wide research evaluation programme[15], we contribute to that by unfolding the patterns in climate innovation overtime. We reveal how research impacts on actionable knowledge using an AI-enabled, cost-effective and scalable methodology. We survey the European innovation landscape by using a database of projects funded between 2002 and 2020 under the last three European Framework programme (FP): FP6, FP7 and Horizon 2020 (H2020). We link projects and peer-review literature by applying Natural language Processing (NLP) to the corpora of abstracts on both layers and by investigating links between disciplines through a network of interconnected topics. The analysis of the network informs on whether links exist between different knowledge branches and reveals the most central topics. We finally compute the lexical distance between topics of the two domains: research (literature) and action (projects). The lexical distance is computed at topic level, expressed by bag of words. We do that to inform European policy about existing gaps and missed opportunities.

We contribute to both innovation studies and evidence-based policy by: (i) first, we quantitatively and qualitatively map the scientific knowledge produced in climate change and we follow its evolution over time; (ii) secondly, we compare and link this knowledge with the projects delivered in Europe to stress potential synergies and bottlenecks. From a policy perspective, the paper opens a science-driven ground to improve the mission-oriented innovation policy framework launched by the European Commission. Finally, the paper contributes to the existing literature with a cutting-edge methodology, which combines machine learning and network science in a unified framework.

**Results**

**The state of research on climate innovation**

We survey a universe of 5556 peer-review articles as derived from bibliometric databases and published between 1979 and 2021 (Table 1, Appendix A). Interest in climate change innovation grew at quasi exponential rate since 2006. This is partly driven by the general increase in academic production that affected almost any research discipline[16], but also the increasing awareness on climate change causes and the need to find solutions. After the Kyoto protocol signature (1999), research embraced a solution-oriented approach which opened to new scientific disciplines. Social scientists entered the climate debate also thanks to the publication of The Stern Review (October 2006) and the momentum expanded further after the Paris Agreement in 2015, when countries strongly committed to keep the temperature "well below 2°C and to pursue efforts to limit the temperature increase to 1.5°C above pre-industrial levels"[17]. These pivotal events also justify the large interest we observe in technology and the strong call for emission reduction and energy transition.

We investigate the heterogeneity of words used (*lexical diversity*) by splitting our database using the IPCC Assessment Reports (ARs) as key reference periods (Table 2, Appendix A). We include 2021 publications up to the release of Working Group I AR6 (August 2021). The lexical diversity signals if new concepts – expressed as new terms – enter the debate. We find that diversity increases overtime (Figure 1) especially with the publication of AR5 (2014). AR5 first introduces the need the 1.5C target, it explores pathways to reach it and provided a detailed overview on the role of technologies (including negative emission technologies, NETs) in advancing solutions to climate change. Findings were scattered in different sections. It was only with the IPCC Special Report on Global Warming of 1.5C (2018) that this knowledge was gathered together. The report advocated the NETs – defined in the report as "carbon dioxide removal (CDR) technologies – "on the order of 100-1,000GtCO2 [billion tons] over the 21st century." [18].

The heterogenous evolution is well captured by the most frequently used words across different IPCC ARs. Prior to the 1990s (AR1), climate innovation research puts strong emphasis on the scientific background and knowledge about climate change impacts on biochemical processes and ocean dynamics. Innovation takes the form of experimental approaches that enhance problem conceptualization and framing. The second IPCC AR timeframe shifts the attention towards energy generation and use. Peer review articles published within this timeframe well capture this trend.

The most locally cited record in this timeframe[19] explores energy solutions of the future accounting for their social, technical and economic constraints and opportunities. Between 1995 and 2000 (AR3 timeframe), economics enters the debate.

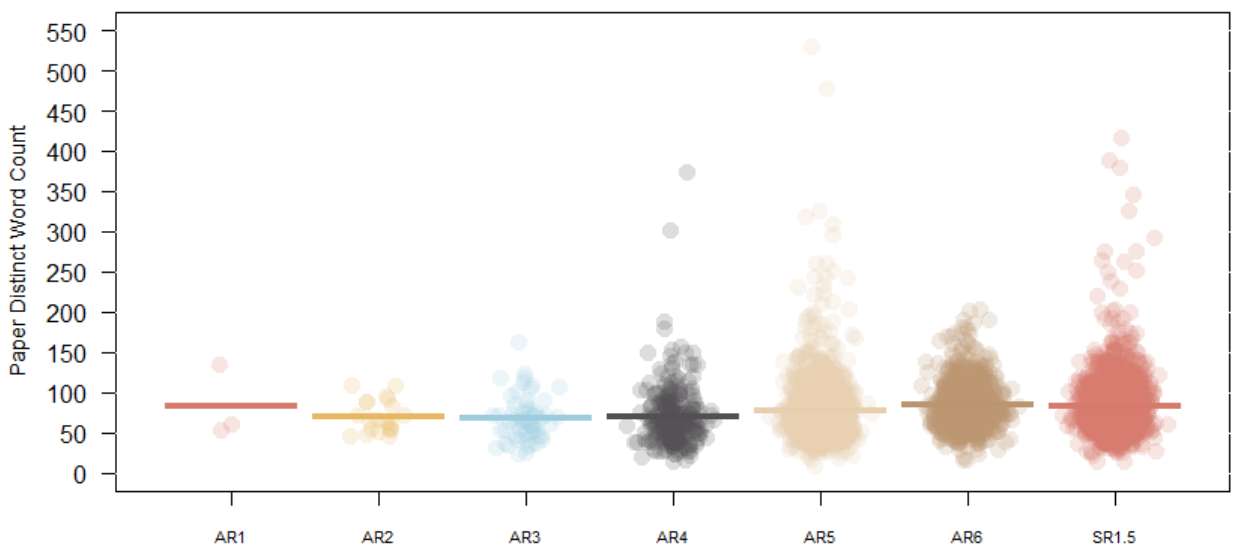

Figure 1 | The lexical diversity of peer-reviewed articles per IPCC Ars. Lexical diversity increases as new terms (expressed as unique words ever appeared before) appear. Lexical diversity is a proxy measure of interdisciplinarity: jargon in climate innovation literature evolves and new disciplines appear.

Among the top cited in this period, some papers tackle the research frontier in climate economics and policies[20–22], while others explore the challenges in modeling technological change[23,24]. Policy innovation become the priority focus for the following six years (AR4): designing the right policies to promote the energy transition[25], encouraging the best response from businesses[26,27] and improving the modeling approaches to technological change[28–30] become priorities for the research community. On the other hand, science progress in addressing the need for quality-assured climate forecasts in water-resource management[31], advancing the atmospheric-data assimilation processes[32] and exploring the connections between the present and the climate of the past[33]. The entrance in the new decade (2007-2013, AR5) is characterized by the role of technology but with a significant shift in sector-specific requirements: nanotechnologies in water usage and reuse[34], transports[35] and rain-fed agriculture[36]. There is a strong methodological focus on innovation systems[37], community-based approaches[38] and complexity[39]. Research published from 2015 (SR1.5 and the then in progress AR6) looks at the transition towards a climate-neutral economy by tackling innovative energy production processes[40–44] and by exploring the role of essential climate variables in policies and applications[45]. This higher specialization of topics calls for a strong reflection on interdisciplinarity, but also signals how specific science-based solutions can be.

From the application of the Structural Topic Modelling (STM) a sub-discipline of Natural Language Processing (see Methods section), we find 34 topics (Table 3, Figure 1, Appendix A), which describe the bulk of climate-change innovation as tackled and explored by research. The topics are clustered into three broad dimensions: adaptation, mitigation and product innovation with high and low technological intensity.

We observe a growing interest (Figure 2, Appendix A) in adaptation tools and practices, which include blue and green infrastructure (Topic 15), improved water resource management practices and tools (Topic 16), climate-smart genetically modified crops (Topic 25) and novel assessment methods (Topic 32). On the other hand, mitigation-related topics exhibit a more stable trend except for innovations targeting vulnerable and developing countries (Topic 10), low-carbon technologies in China (Topic 19) and energy efficiency technologies (Topic 29). Research has increasingly questioned the governance structure of climate change (Topic 27) and has been progressively suggesting new and cost-effective ways to investigate the economics of technological change (Topic 30). We also observe a steady growth of research published around biomass and bioproducts (Topic 22) and a sustained interest for climate services (Topic 33) and health-related risk mitigation tools (Topic 9). This landscape suggests a progressive interdisciplinary attitude towards innovation, but also launches call for action to specific disciplines: there is wide room for economists (Topic 18) to tackle viable solutions while meeting the society's needs confirming Vinuesa at al. (2020)[3].

Beyond the simple investigation of core topics in the research domain, we are interested in mapping their connections. Topics devoted to governance and process innovation are directly linked to climate services and tools which require interdisciplinary competences to come together (Figure 3). Adaptation and information services call for strong co-generation processes, which may result in novel structures and new actors involved. We also find that the topics related to energy-intensive sectors (i.e. transport and electricity production) are linked to each other and signal the need for business innovation and large-scale diffusion of technologies. The third highly-connected group highlights that researchers pay attention to the complex human-nature interaction when promoting solutions to fight climate change. It is the case of nature-based solutions, but also to climate-smart agriculture and blue and green infrastructure.

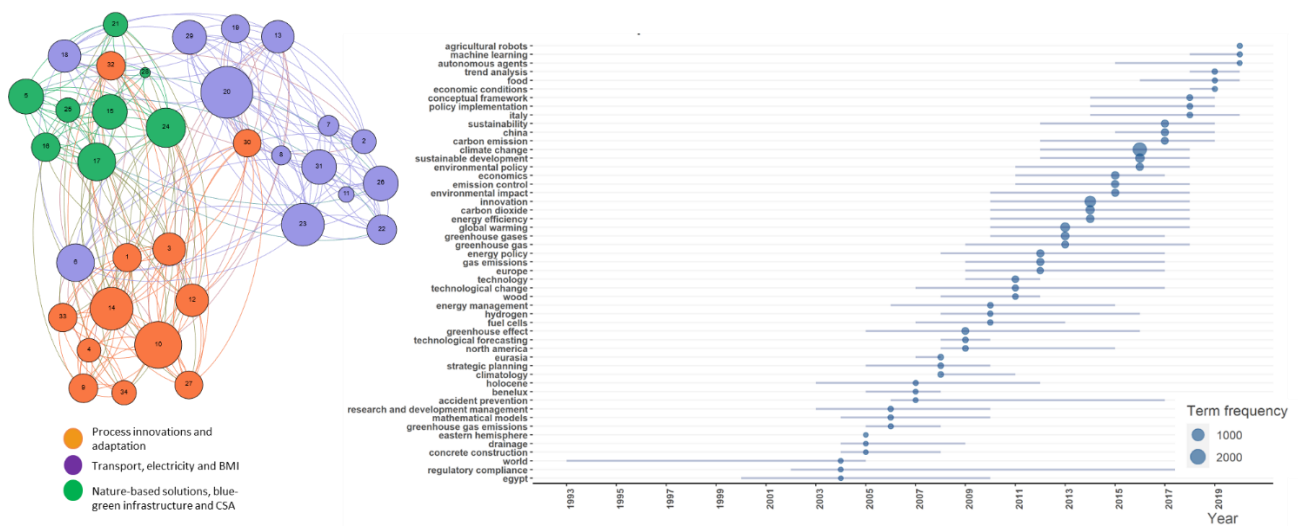

Figure 3 | The topic correlation in the research layer. (Left) Clusters of topics and their interactions, where node size reflects the prevalence (in-degree) of topics. Adaptation is strongly linked to interdisciplinary and information-based services (orange bubbles); mitigation (purple bubbles) is primarily described by energy-intensive sectors, such as transport and electricity production; green bubbles represent topics that span across the mitigation and adaptation domains. (Right) The most frequent terms in the corpus of documents and their evolution overtime

**Assessing EU-action priorities**

We analyse the abstracts of 2067 projects funded between the 2004-2020 timeframe and articulated as follows: 198 under FP6, 918 under FP7 and 951 under H2020. Following the same methodology presented for scientific papers (see Methods), we find 33 core topics describing the corpus of EU-funded projects (Table 4, Figure 4, Appendix A). Among them, the best represented topics relate to the next generation of Earth monitoring tools, climate services and nature-based. We observe a growing number of projects involved in technological development, including progresses in the transport industry (Topic 11), improved ways to generate and distribute electricity (Topic 16), new materials to achieve energy efficiency (Topic 20). Adaptation-related innovations gained popularity with time. This is the case for energy efficiency and circular economy solutions in urban areas (Topic 13), climate smart crops and transformations in agriculture (Topic 18), nature-based solutions (Topic 26) and risk mitigation innovations in case of flood events (Topic 25). We also observe a growing number of projects related to bioenergy and agrifood solutions (Topic 29) as well as raising interest in biodiversity impact assessments through data-driven applications (Topic 33). Innovations in mitigation technologies are stabilized over the timeframe, with exception for negative emission technologies and carbon capture and storage innovations. This trend suggests that some applications reached a maturity stage and should be scaled up for market.

We map the links between different topics mirroring the procedure we used for the corpus of peer-reviewed publications. We find four groups that represent meta-topics each describing a different dimension of innovation actions (Figure 4): adaptation, mitigation, applied (technical) and theoretical innovation. Within the adaptation domain, we find solutions that translate climate information into usable knowledge. Climate services (Topic 3), early-warning systems (Topic 5) and flood risk assessment tools (Topic 25) fall in this group. The prevalence of these topics collective increase overtime mirroring the growing demand for adaptation especially in most affected areas. Technical applications in mitigation are also increasingly important with market ready energy efficiency measures and innovation in the agroforestry sector (*i.e,* carbon capture and storage) as key areas. On the theoretical side of mitigation, topic prevalence is more stable as advancements in these space are more consolidated: progresses in economics and policy (Topic 7) and models and tools to improve the understanding of water cycles (Topic 4) are the top ranked. Finally, theoretical innovations in the adaptation sphere are mostly influenced by the high degree of interdisciplinary they are built on. Here, we recognize new methods to biodiversity protection and preservation (Topic 10), as well as novel communication approaches which include co-production and stakeholder involvement (Topic 1).

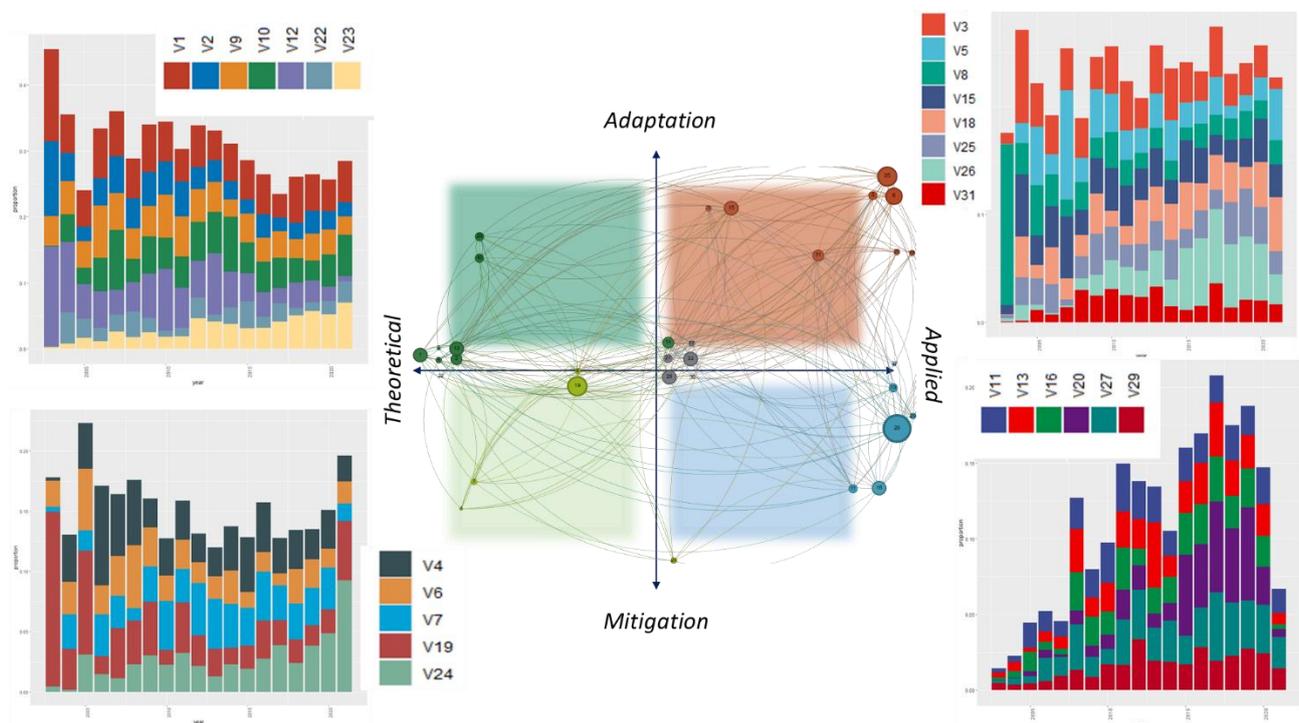

Figure 4 | The topic correlation in the project universe and their dynamic evolution. The four clusters of topics are represented as a network on a four-dimensional plane. The size of each node represents the in-degree centrality: how many links each topic receives. It is a proxy for relevance of every given topic. The top topics by centrality are bioenergy and agri-food solutions (Topic 29), air and water quality monitoring (Topic 19) and flood risk assessment tools (Topic 25). The dynamic evolution of each area is displayed in the staked barcharts with prevalence (i.e., relevance of each topic over the whole sample) on the y-axis. The full list of topics is included in the Supplementary Material.

**Measuring the research-action distance**

We measure the density of the networks of research's and project's topics to capture hoe many relationships between actors exist compared to the all possible ones. The network of research topics is denser (0.2245) than the projects' one (0.1988), confirming there is stronger contamination between different areas of work in research[2] that can become an advantage for projects, too. We find significant differences in the composition of the topics clusters (Figure 6, Appendix A). First, economic incentives and price mechanisms are strongly bonded to energy efficiency and low-carbon technologies in the research domain. Hence, they are seen as enablers of large-scale diffusion of mitigation practices. In the project domain, instead, economic instruments, models and theories are associated to transformation pathways. They serve to promote adaptation practices, to value nature-based solutions and to improve energy efficiency in cities.

Agriculture marks the second biggest discrepancy between the two domains. In the corpus of peer-reviewed publications, the agri-food sector mainly interacts with adaptation-related topics. The cosine similarity reveals that improved water resource management and smart solutions to climate-related variables are quite related to agriculture (similarity equal to 0.40 and 0.34 respectively). Agriculture, biomass and bioenergy production are less similar (0.20). On the contrary, projects tackle agriculture to reduce its impact on the environment. Climate-resistant crops and transformations in agriculture are addressed in tandem with biodiversity preservation (cosine similarity = 0.73) and freshwater ecosystem (cosine similarity = 0.38).

The third remarkable difference relates to innovations in industrial processes and new materials. The peer-reviewed documents investigate opportunities stemming from life-cycle assessment methods and focused on evaluating the environmental performance in energy-intensive industries, such as transport and electricity production. The topics in the projects' set cover innovations and solutions to reduce emission in other industries, such as aviation and agriculture. Both domains include energy efficiency improvements, but the projects' one stresses the market readiness of these solutions.

Our cosine-similarity-based distance measure provides an understanding of where major research-action gaps are. While domain-specific connections are comparable, some topics lack the adequate inputs for research and/or implementation in innovation actions (Figure 5). Research progresses in biomass and bioproducts (R22) and in genetically modified crops (R25) have still few operational applications. The former includes the forest products and biorefining industry, where innovation is crucial to reduce the environmental footprint of both products and processes. The latter entails innovation in agriculture and specifically in biotechnology-led approaches to uplift agricultural production, while also questioning existing methods to support climate-stressed developing countries. Our results also call for wider operationalized knowledge in risk mitigation and adaptation practices failed to find adequate application (R17). This research topic strongly looks at emerging economies and vulnerable areas of the world. Testing research advancements and recommendations in the real world is more urgent than ever. We also find lose connections in material innovations including critical minerals and extraction of primary resources to support the energy transition (R11). Finally, product innovation and corporate social responsibility (CSR) seems also not adequately connected to operationalized knowledge. CSR includes labelling and certifications that guide consumers towards green or climate-smart products. Also, it includes sustainability practices to reduce corporate carbon footprint.

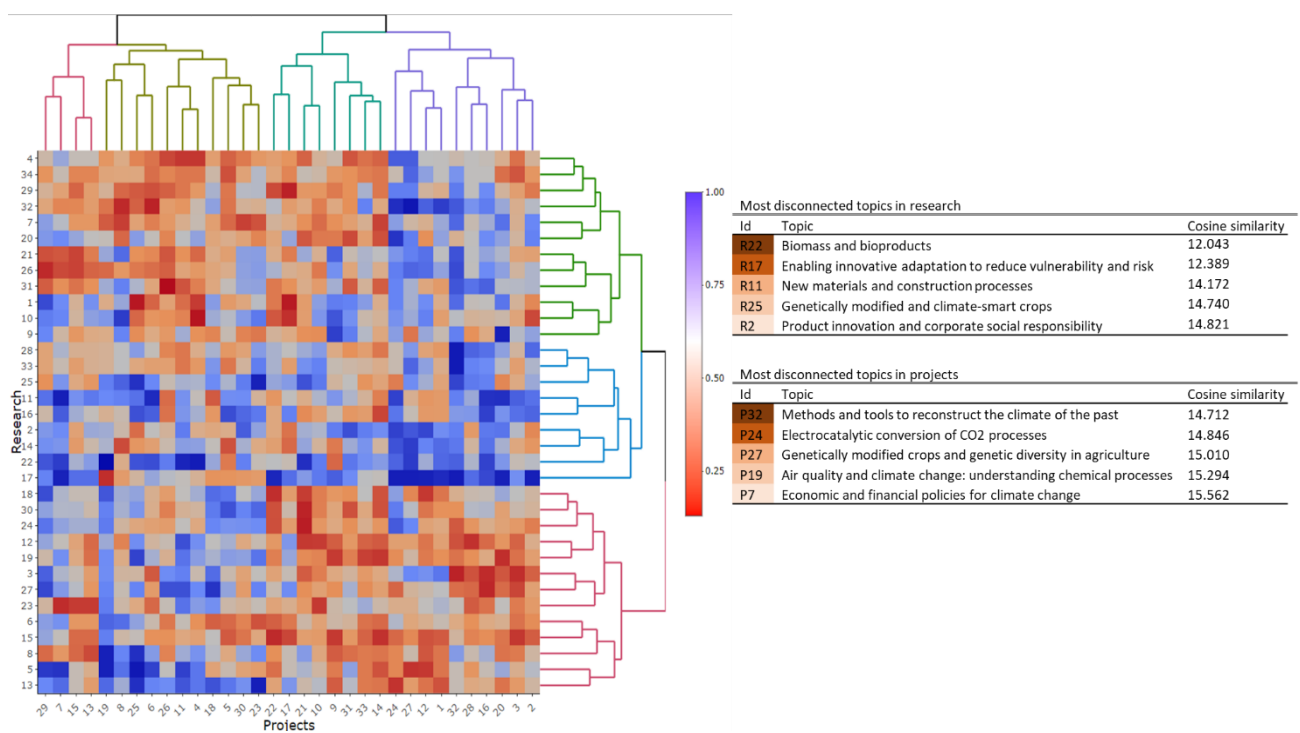

Figure 5 | Cosine similarity ranges between 0.13 and 1 implying that every topic has at least some connections. The tables display cumulative cosine similarity values. In the research domain, adaptation topics are the ones suffering the most from lose connections with applications. Exceptions are new materials (including critical minerals) and CSR. On the projects side, both technical (mostly chemical processes for past and present climate models) and social (economics and policy) aspects should receive higher attention.

On the projects' domain, research is needed to expand both technical and socio-economic understanding of causes and impacts of climate change. Three out of the top five least connected topics entail chemical and technical processes to tackle CO2 concentrations. The first belongs to the domain of paleoclimatology: while projects run, research is slower in showcasing results. Innovation here lies in methods, including numerical models and sampling techniques. One of the reasons behind this result is the resource intensity of the disciplines: the reconstruction of the climate of the past often involves long and intense expeditions in remote regions and the coordination of large, multidisciplinary teams. Biases in our dataset also explain the cosine similarity score: our query may partly rather than fully capture paleoclimatology, limiting the recognition of the discipline. Other technical topics would benefit from research effort to evaluate actions, highlight their benefits, and unleash their barriers. It is the case of engineer-based processes to recover and convert greenhouse gas emissions (P24). Projects develop solutions for carbon capture and storage, electrocatalytic conversion of $CO_2$ into chemical energy carriers, as well as new CO2 capture processes by innovative absorbents based on novel aerogels. All these products must be further studied to understand their climate and socio-economic potential, as well as to test their usability in diverse. We mark the same needs for genetically modified crops (P27), which may suffer from barriers that limit their uptake: research would help with their identification.

**Discussion and conclusions**

The European Green Deal calls for fostered "deployment of innovative technologies and infrastructures" (Section 2.1.2., [46]) and the US Green New Deal – without explicitly mentioning the term "innovation" lists the widespread needs top "invest in the infrastructure and industry of the United States to sustainably meet the challenges of the 21$^{st}$ century" (p. 5) [47].). More recently, RePowerEU reinforced the need to expand R&D activities to accelerate the transformation of our energy system and to achieve the European Green Deal objectives.

Innovation is a process that leads to a "creative destruction"[48] capable of transforming the way the productive means works. Innovation is a chain that takes "time and money"[49]. It begins with invention and it develops with R&D, failures and tests. Research does not just provide advancements in methods, but also opens to new stakeholders, approaches, and needs. As the disruptions in human and natural systems caused by climate change are "unequivocal[50], the establishment of a productive dialogue between research and practices has never been more urgent.

In this work, we measure the research-action gap in climate innovation. We explore to what extent and how the academic progresses – as described by the corpus of peer-reviewed literature in the field – is perceived and applied in EU-funded innovation projects. The focus on the EU innovation actions has a double motivation: the mission-oriented European framework aims at shifting from useful to usable science to achieve the milestone missions of at least the next framework. Moreover, the open data policy of the European Union provides us with the rare opportunity of a wide database of detailed projects. We leave the research layer of our investigation open to extra European contributions as published records do not know boundaries and their exploitation is not restricted to specific geographical areas.

While our approach contributes to the existing literature in innovation studies, our results can inform the European climate-change policy. We find a sustained growing interest in adaptation measures and processes and a stabilized landscape of mitigation-oriented innovation. This trend may suggest that at least part of the low-carbon energy technologies developed in the past are now mature for their reference market [51]. Exceptions exist and include new computational methods (*i.e.,* machine learning and AI) and novel technologies (*i.e.,* carbon capture and storage). The raising focus on adaptation and risk mitigation solutions flags that impacts have been and will continue to be felt across the world.

Among our findings, two call for stronger attention from policy making. First, the research and action layers differ in the way different topics link to each other. For instance, research has explored economic tools (*i.e.* prices and incentives) mostly in connection to energy efficiency and low-carbon technologies, while much has to be done in the adaptation sphere. This includes economic assessments and evaluation of existing projects from a scientific perspective. Evidence in this sense would boost the diffusion of good practices showing what works and in which contexts. Similarly, projects can make wider use of interdisciplinary research. This holds for agriculture and improved use of natural resources, where research has important lessons learnt still poorly explored by EU projects.

Second, the distance between research and action reveals that there is room for stronger interactions between some key areas: large-scale agriculture, industrial applications and risk assessments. A tighter collaboration with the industry and private players could play a role in advancing the uptake and diffusion of climate-smart innovations. In agriculture, climate-resistant crops can guarantee food security and increased resilience to the most vulnerable. A stronger cooperation between researchers, civil society and decision-takers could boost the development and test of these applications while also protecting biodiversity and limiting the negative environmental consequences of their exploitation. We find that the distance is mostly due to projects' limits, hence suggesting that some of the research findings could be better incorporated into the actions.

As for industrial applications, we find a significant research-action gap in bioproducts. Here, a tighter cooperation with industrial players and the private sector would lead to standards, certificates and shared practices that projects often develop with insufficient support from research. This holds for business model and market innovation. Research and action can further cooperate over risk assessment and especially flood risks. Here, projects can serve to collect granular exposure and vulnerability data serving as seeds for punctual and precise models. Basic and applied research should be supported to deliver actionable knowledge and to increase the accuracy of climate projections and forecasts. On the other hand, projects can act as interfaces with stakeholders to promote a climate-aware and risk-conscious culture.

The methods here proposed provide a tool for policymakers to understand how research and innovation are intertwined, and to select specific actions to address slow progress in priority areas. For researchers, these findings can support identifying research and innovation gaps to address. As the old adage goes, you cannot achieve what you cannot measure. Understanding and continuous monitoring of the research, innovation and action gaps is crucial for holistic climate action.

**Methods**

We retrieve data from two different sources. For the peer-reviewed literature, we use the bibliometric database Scopus. We build our corpus of publications by launching three large and combined queries that capture the historical evolution of concepts related to climate change (details are provided in the Appendix A) on Scopus in November 2020. The three expressions capture: the jargon changed as the science progressed. The query embraces the shift from a pure "warming" idea to a more complex and transdisciplinary one: "change".

For the EU-funded innovation actions, we downloaded EU-funded projects under three frameworks: FP6, FP7 and H2020 from the European Commission Community Research and Development Information Service (CORDIS) database (https://cordis.europa.eu/en). The database contains information about the name and the id of the project, call, the program, the project's cost and the

European maximum contribution to it, the coordinator's name and country as well as the participants to the consortium. Importantly for us, it contains the abstract of the project in the form of short text. We filter only climate innovation relevant projects by removing other areas of interest and by applying the same large query used for peer-reviewed literature.

We apply an unsupervised machine-learning routine to both the abstracts of the research papers and innovation actions' ones. Written text offers valuable information about the linguistic, semantic and contextual nature of documents. The growing amount of literature, reports and written records in the space of climate action and climate innovation, complicates the process of turning information into usable knowledge. A convenient approach for the analysis of large text corpora stems in the use of topic models [52] or mixed membership models [53]. Topics are semantic "themes" generated by distributions of words belonging to a specific vocabulary. In mixed membership models, each document is described by multiple topics; the words in documents are assigned to single topics. Hence, documents are represented as vectors of words assigned to different topics according to an estimated proportion. Language is an information-dense mode of communication for humans, but the alphabetic characters which comprise words have no intrinsic quantitative value for computation. Statistical language modeling[54] can be used to train on a body of language and embed the meaning of every word in a vocabulary into a machine-readable vector. This vector is a quantitative representation of semantic space such that spatial metrics such as cosine distance can be used to assess similarity between words; for example the vector ("Man") subtracted from the vector ("King") added to the vector ("Woman") results in a vector which is closest to that of ("Queen") [55]. Such high-fidelity representations are a useful tool for estimating semantic similarity to process the large scale of growing climate-related documentation.

We use a structural-topic model (STM) to first detect the most relevant topics and then to quantify their potential connections. The STM [56] is an improvement of multiple topic-modelling techniques including the latent Dirichlet allocation (LDA) [57], the correlated-topic model (CTM) [58] and multiple extensions of these two [59–61]. The novelty of STM lies in its ability to discover the topical structure of the corpus estimating the relation of the topics with the documents' metadata (e.g. publication years, framework programs). As other topic models, STM is an "unsupervised" and data-driven method. STM does not ex-ante assume the optimal number of topics, nor their content. Instead, it infers them from data. Differently from standard LDA, STM achieves a better distribution of words in topics by exploiting the covariance with document-specific metadata.

The process is explained as follows: each document is a set of $K$ topics, as in the standard LDA model; we estimate a distribution over topics ($\theta$) to obtain topic proportions, and the topics can be correlated with each other, where their prevalence may vary because of a set of covariates $W$. This is particularly true in presence of a highly interdisciplinary field such as climate change innovation. The prevalence is estimated through a standard regression model, where $W \sim LogisticNormal(W\gamma, \Sigma)$ and $\Sigma$ is a $K-1$-by-$K-1$ covariance matrix. First, topic proportions are estimated by assigning words to the document-specific distribution of topics conditional to covariates $W$. For instance, we may be interested in topics proportion depending on the framework programme under which projects are funded. Words are then chosen from a multinomial distribution with parameter $\beta$, conditional to the topic assignment in the first step. The parameter $\beta$ derives from deviations from the baseline of words distribution, which can vary because of a set of covariates $U \neq W$. For instance, we may find that topics related to "adaptation" to climate change" use frequently words like "resilience".

The STM has three core differences with respect to standard LDA. The first is the potential correlation between topics; the second is the document-specific prior distribution over topics; the third is the presence of an additional set of covariates within topics $U \neq W$. The STM is a replicable and time-effective method with few a priori assumptions to be made. However, it requires a wide effort to interpret the topics as bags with word- and document-probabilities assigned.

We include different covariates per layer. In the peer-reviewed research layer, we split the database according to the different IPCC Assessment Report (AR), a series of six-year reviews of the existing knowledge and science on climate change, (Table 1, Appendix A). We are interested in understanding whether the timeframe to which the publication belongs influences to any extent the prevalence of some topics over others. As second set of covariates, we use the publication year to allow for annual heterogeneity within the sample. As for the set of EU-funded projects, we let the topics prevalence vary per framework programme under which projects receive funds (i.e. FP6, FP7 and Horizon2020).

The STM allows the identification of the topics characterizing both layers (i.e. peer-reviewed publications and EU-funded projects) and their connections. We are not only interested in the intra-layer connections, but also in the distance between the research layer and the project one. Multiple areas of enquiry follow a complex dynamic. On one hand, topics can lead to aggregate phenomena that signal the strong interest of one layer towards new emerging properties. On the other, missed links and loosely connected topics highlight where efforts should be concentrated to strengthen the research-action collaboration. First, we use the STM to compute links between topics and generate a network of interaction weighted on their correlation coefficient. We repeat the process for both layers separately. Then, we measure the distance between the two networks by computing the cosine similarity between topics belonging to the two layers. This approach measures how similar two pieces of text are based on their use of words sharing the same vocabulary (English in this case). The cosine similarity is a metric computed between documents (say A and B) in an n-dimensional vector space, where every word is a vector:

$$\cos(\theta) = \frac{\sum_{i=1}^{n} A_i B_i}{\sqrt{\sum_{i=1}^{n} A_i^2} \sqrt{\sum_{i=1}^{n} B_i^2}}$$

This indicator ranges from 0 to 1, and the measurement of the distance between the layers and their connections brings new insights into the literature. On the one hand, it offers a fast and effective way to estimate the research-action gap. This approach could potentially be applied to other domains of enquiry, beyond climate change innovation. On the other hand, we observe how connections and links between diverse topics fill existing gaps. We observe the collective emerging properties of the networks and the difference between the two layers under scrutiny. Doing so, we achieve a comprehensive view of climate-change innovation and contributions to enable a stronger interaction between theory and action.